\documentstyle[12pt,cite,amssymb,epsf]{article} 

\title{
\begin{flushright}
{\normalsize Yaroslavl State University\\
             Preprint YARU-HE-00/03\\
             hep-ph/0003216} \\[5mm]
\end{flushright}
Lepton pair production by high-energy neutrino 
in an external electromagnetic field}

\author{{A.V.~Kuznetsov, N.V.~Mikheev, D.A.~Rumyantsev}\\ [7mm] 
{\small\it Division of Theoretical Physics, Department of Physics,}\\
{\small\it Yaroslavl State University, Sovietskaya 14,}\\
{\small\it 150000 Yaroslavl, Russian Federation}}

\date{}

\begin{document}

\maketitle

\begin{abstract} 
The process of the lepton pair production by a neutrino propagating in an 
external electromagnetic field is investigated in the framework of 
the Standard Model. Relatively simple exact expression for the 
probability as the single integral is obtained, which is suitable 
for a quantitative analysis. 
\end{abstract}

Nowadays, it is well established that a medium makes an active influence 
on the quantum processes. It stimulates a constantly growing interest 
in the particle physics in medium, especially in view of possible 
astrophysical manifestations.
It should be noted that a consideration of an intense electromagnetic field
as the medium, along with a dense matter, is physically justified indeed. 
Really, the field 
strengths inside the astrophysical objects can reach the critical 
Schwinger value $B_e = m_e^2/e \simeq 4.41 \cdot 10^{13}$ G
\footnote{We use natural units in which $c = \hbar = 1.$}, 
and even could exceed it essentially. 
On the other hand, the situation is possible when the so-called field 
dynamical parameter $\chi$ of the relativistic particle~\footnote{Its 
definition is:
$\chi = e (p_\alpha F_{\alpha \beta} F_{\beta \sigma} p_\sigma)^{1/2}/m^3$, 
where $p_\alpha$ is the particle four-momentum, $F_{\alpha \beta}$ is the 
electromagnetic field tensor.} 
propagating in a relatively weak electromagnetic field, $F < B_e$ 
($F = {\cal E}$ and/or $B$), could appear rather high. 
In this case the field in the particle rest 
frame can exceed essentially the critical value and is very close to 
the crossed field ($\vec {\cal E} \perp \vec B$, ${\cal E}=B$). 
Thus, the calculation in a constant crossed field
is the relativistic limit
of the calculation in an arbitrary relatively weak smooth field. 
Consequently, the results obtained in a crossed field 
possesses a great extent of generality, and acquires interest by itself.

It is known that such intense electromagnetic fields 
allow the processes which are kinematically 
forbidden in vacuum, such as the creation of the lepton pair by a neutrino, 
$\nu \to \nu \ell^- \ell^+ \, (\ell = e, \mu, \tau)$. 
It should be noted, that the $\ell^- \ell^+$ pair 
can have a sufficiently large space-like total momentum
in an electromagnetic field, due to the specific kinematics of 
a charged particle in this field. 
Therefore, the process with a relativistic neutrino becomes purely 
diagonal with respect to the neutrino flavor and is insensitive to its mass 
and to the mixing in the lepton sector.

The theoretical study of the process of the electron - positron pair 
production by a neutrino in the crossed field limit has a rather long 
history~\cite{Chob,Bor83,Knizh,Bor93,KM97}.
A correct type of the dependence of the probability 
on the dynamical parameter $\chi$ in the leading log approximation, namely, 
$\sim \chi^2 \ln \chi$, was found in the paper~\cite{Chob}, where the 
numerical coefficient was wrong, however. 
In the succeeding papers attempts were made to adjust this coefficient and 
to find the next post-logarithm terms which could appear quite essential 
when $\ln \chi$ was not very large. 

According to the definition of the problem in the crossed field 
approximation, one should speak about the 
ultrarelativistic neutrino only, which exists as the left-handed one due 
to the chiral type of its interaction in the frame of the Standard Model, 
even if the neutrino mass is non-zero. 
A lack of understanding of the fact that nonpolarized ultrarelativistic 
neutrinos did not exist in Nature, often led to the appearance of an 
erroneous extra factor 1/2 in expressions for the probabilities of the 
processes with initial neutrinos, due to an unphysical averaging over the 
neutrino polarizations, see e.g. the papers~\cite{Skob95,Bor99}. 

The results for the probability of the process $\nu \to \nu e^- e^+$ 
in the crossed field, which were obtained in the listed papers, had 
essential distinctions. 
This probability for the case of a high-energy neutrino ($\chi \gg 1$) 
can be presented in the following form 
\begin{equation}
w(\nu \to \nu e^- e^+) \, = \, K \, w_0 \,\chi^2 
\, \left(\ln \chi - {1 \over 2} ln 3 - \gamma_E + \Delta \right),
\label{eq:w0}
\end{equation}

\noindent 
where
\begin{equation}
w_0 \, = \, 
\frac{G_F^2 \,(g_V^2 + g_A^2) \,m_e^6}
{27 \pi^3 E_\nu},
\label{eq:w00}
\end{equation}

\noindent 
$\gamma_E = 0.577 \dots$ is the Euler constant, $g_V, g_A$ are the constants 
in the effective local Lagrangian of the $\nu \nu e e$ interaction, see 
Eq.~(\ref{eq:L}) below. 
In the recent paper~\cite{Bor99} devoted to the 
study of the massive neutrino decay $\nu_i \to \nu_j e^- e^+ 
\,(m_i > m_j + 2 m_e)$ in an external field, a comparance was also made 
of various results for the process probability. However, the statement which
was made in~\cite{Bor99} about a mutual agreement of the results was 
incorrect, in our opinion. 
Really, the constants $K$ and $\Delta$ introduced in 
Eq.~(\ref{eq:w0}), were obtained by the authors as follows, see Table 1.

\begin{table}[h]
\caption{The constants $K$ and $\Delta$ in Eq.~(\ref{eq:w0}), 
obtained in various papers}

\begin{center}
\begin{tabular}{|ll|c|c|}\hline
& & &  \\ 
& & $K$ & $\Delta$ \\[5mm] 
\hline 
& & &  \\ 
Choban, Ivanov & 1969~\cite{Chob}
& $\frac{29}{1024 \pi}$ & --- \\[5mm]
Borisov et al. & 1983~\cite{Bor83}
& 1 & $- 2 \ln 2 - {389 \over 384} + {9 \over 128} 
\frac{g_V^2 - g_A^2}{g_V^2 + g_A^2}$\\[5mm]
Knizhnikov et al. & 1984~\cite{Knizh}
& ${9 \over 16}{E_\nu \over m_e}$ & --- \\[5mm]
Borisov et al. & 1993~\cite{Bor93}
& ${1 \over 2}$ & $ + {5 \over 4} $\\[5mm]
Kuznetsov, Mikheev & 1997~\cite{KM97}
& 1 & $ - {29 \over 24}$\\[5mm]
Borisov, Zamorin & 1999~\cite{Bor99}
& ${1 \over 2}$ & $ - {29 \over 24}$\\[5mm]
\hline
\end{tabular}
\end{center}

\end{table}

Note that in the papers~\cite{Chob,Bor99} calculations were performed 
for the case 
of the electron -- neutrino interaction via the $W$ - boson only. 
To compare our Eq.~(\ref{eq:w0}) with the results of these papers,
one should suppose in this formula $g_V = g_A = 1$~\cite{Chob}, 
and $g_V = g_A = |U_{ei} U_{e3}|$~\cite{Bor99}, correspondingly. 
Loss of the factor $m_e/E_\nu$ in the resulting expressions for the 
probability in Ref.~\cite{Knizh} was not the numerical mistake but 
rather the physical one, because it led to the loss of Lorentz
invariance of the product of the probability and the neutrino energy.

As it was mentioned above, the formula~(\ref{eq:w0}) for the probability 
described a rather particular case of $\ln \chi \gg 1$. 
On the other hand, the situation is realized under some physical conditions 
where the dynamical parameter takes the values which are moderately large, 
namely, $\chi \gg 1$, but $\ln \chi \sim 1$. 
The crossed-field approximation is valid in this case, but the condition
$\ln \chi \gg 1$ fails, and a consideration is necesary in Eq.~(\ref{eq:w0}) 
of the next terms of expansion over the inverse powers of the large 
dynamical parameter $\chi$. The expressions for the probability at an
arbitrary value of the $\chi$ parameter, presented in some of the listed 
papers, have a cumbersome form of multiple integrals, which are not suitable 
for an analysis. 

Hence the problem is urgent of obtaining the probability of the lepton pair 
($e^- e^+$ or $\mu^- \mu^+$) production by a neutrino propagating in 
the crossed field for an arbitrary value of the $\chi$ parameter. 
In this paper we present our result for the probability of the process 
$\nu \to \nu \ell^- \ell^+$ which is rather simple and suitable for 
a quantitative analysis. 

We will consider the case of relatively low momentum transfers
($|q^2| \ll m_W^2\,$). Under this condition, the weak interaction of 
neutrinos with charged leptons can be considered in the local limit 
by using the effective Lagrangian 

\begin{equation}
{\cal L} \, = \, \frac{G_F}{\sqrt 2} 
\big [ \bar \ell \gamma_{\alpha} (g_V - g_A \gamma_5) \ell \big ] \,
\big [ \bar \nu \gamma^{\alpha} (1 - \gamma_5) \nu \big ] \,,
\label{eq:L}
\end{equation}

\noindent where $g_V = \pm 1/2 + 2 sin^2 \theta_W, \, g_A = \pm 1/2$.
Here the upper signs correspond to the situation when the neutrino flavor 
coinsides with the lepton $\ell$ flavor ($\nu = \nu_\ell$), 
in this case both $Z$ and $W$ boson exchange takes part 
in a process. The lower signs correspond to the case 
$\nu = \nu_{\ell'}, \, \ell' \ne \ell$, when the $Z$ boson exchange 
is only presented in the Lagrangian~(\ref{eq:L}). 

We omit the details of calculations which can be found e.g. in our 
paper~\cite{KM97}, and present here the result for the probability 
in a form of the single integral containing the Airy function:

\begin{eqnarray}
w(\nu \to \nu \ell^- \ell^+) 
 = \, \frac{G_F^2 \,(g_V^2 + g_A^2) \,m_\ell^6 \,\chi_\ell^2}{27 \pi^4 E_\nu} 
\, \int\limits_0^1 u^2 d u \,z\, \Phi(z) 
\Bigg \lbrace {4 \over 1 - u^2} \left( 2 L(u) - {29 \over 24} \right) 
\nonumber \\
- {15 \over 2} L(u) - {47 \over 48} 
+ {1 \over 8} 
\left(1 + (1 - u^2) L(u) \right) \left(33 - {47 \over 4}
(1 - u^2) \right) 
\nonumber \\
+ {9 \over 16} \frac{g_A^2}{g_V^2 + g_A^2} 
\left[48 L(u) + 2 -
\left(1 + (1 - u^2) L(u) \right) \left(28 - 3 (1 - u^2) \right) 
\right] \Bigg \rbrace.
\label{eq:w1}
\end{eqnarray}

\noindent 
Here $\chi_\ell$ is the dynamical parameter of the lepton with the mass
$m_\ell$, 

\noindent 
$\chi_\ell = 
e (p_\alpha F_{\alpha \beta} F_{\beta \sigma} p_\sigma)^{1/2}/m_\ell^3$, 
$\Phi (z)$ is the Airy function
\begin{equation}
\Phi (z) \, = \, \int\limits_0^\infty d t \cos 
\left(z t + {t^3 \over 3} \right), \quad
z \, = \, \left( \frac{4}{\chi_\ell (1 - u^2)} \right)^{2/3}, \quad
L(u) \, = \, {1 \over 2 u} \ln {1 + u \over 1 - u}.
\label{eq:Phi}
\end{equation}

\noindent 
In the case when $\chi_\ell \ll 1$, one immediately obtains from 
Eq.~(\ref{eq:w1}) the formula for the probability containing the 
well-known exponential suppression:
\begin{equation}
w(\chi_\ell \ll 1)
 =  \frac{3 \, \sqrt{6} \, G_F^2 \, m_\ell^6}{(16 \pi)^3 E_\nu} \,
(3 g_V^2 + 13 g_A^2) \,
\chi_\ell^4 \, \exp \left(- {8 \over 3 \chi_\ell} \right),
\label{eq:w2}
\end{equation}

\noindent 
which agrees with corresponding formula of Ref.~\cite{Bor93}. 

In the case when $\chi_\ell \gg 1$, it is easy to obtain from 
Eq.~(\ref{eq:w1}) the formula~(\ref{eq:w0}), where $K = 1,\, 
\Delta = -29/24$, in agreement with the result of Ref.~\cite{KM97}. 
It is not difficult also to find from our Eq.~(\ref{eq:w1}) 
the next term of expansion over the inverse powers of the 
dynamical parameter $\chi_\ell$. One obtains:
\begin{eqnarray}
w(\chi_\ell \gg 1)
& = & \frac{G_F^2 \,(g_V^2 + g_A^2) \,m_\ell^6 \,\chi_\ell^2}{27 \pi^3 E_\nu} 
\, \bigg \lbrace \ln \chi_\ell - {1 \over 2} ln 3 - \gamma_E - {29 \over 24} 
\nonumber \\
& - & {1 \over \chi_\ell^{2/3}} \, {9 \over 56} \,
\frac{3^{1/3} \pi^2}{\left[\Gamma \left( {2 \over 3} \right) \right]^4}
\, \frac{19 g_V^2 - 63 g_A^2}{g_V^2 + g_A^2} 
\bigg \rbrace, 
\label{eq:w3}
\end{eqnarray}

\noindent where $\Gamma (x)$ is the gamma function.

It is seen from Eq.~(\ref{eq:w3}) that the correction term
$\sim \chi_\ell^{-2/3}$ is not universal. It is relatively small and 
negative in the case when the neutrino flavor coinsides with the charged 
lepton flavor $(\nu_e \to \nu_e e^- e^+, \, \nu_\mu \to \nu_\mu \mu^- \mu^+)$. 
When the flavors of the neutrino and of the charged lepton are different
$(\nu_\mu \to \nu_\mu e^- e^+, \, \nu_e \to \nu_e \mu^- \mu^+)$, 
the correction term is positive and relatively large.

The dependence of the probability of the process
$\nu \to \nu \ell^- \ell^+$ on the dynamical parameter $\chi_\ell$ 
in the region where its value is moderately large, is represented 
in Figs. 1 and 2. One can see that the correction term 
$\sim \chi_\ell^{-2/3}$ is more likely to worsen than to improve 
the presentation of the probability in this region. 
As the analysis shows, this term has a sense just 
of the correction for the values $\chi_\ell \gtrsim 10^5$ only. 

Therefore, our exact formula~(\ref{eq:w1}) should be used in a detailed 
analysis of the probability of the lepton pair production by a neutrino 
propagating in an external electromagnetic field, when the value of
the dynamical parameter $\chi_\ell$ is moderately large. 

\bigskip

\noindent 
{\bf Acknowledgements}  

We thank A.V.~Borisov for useful discussion which stimulated this research.

The work is supported in part by 
the Russian Foundation for Basic Research under the Grant N~98-02-16694.

\bigskip

\centerline{\bf Figure captions}

\bigskip

{\bf Fig. 1}

The dependence of the probability of the process
$\nu \to \nu \ell^- \ell^+$ on the moderately large
dynamical parameter $\chi_\ell$ in the case when the neutrino flavor 
coinsides with the charged lepton flavor 
$(\nu_e \to \nu_e e^- e^+, \, \dots)$:
{\it a)} from the exact formula~(\ref{eq:w1});
{\it b)} from the approximate formula~(\ref{eq:w0});
{\it c)} from the formula~(\ref{eq:w3}) with the 'correction'
$\sim \chi_\ell^{-2/3}$.

\bigskip

{\bf Fig. 2}

The same as in Fig. 1, in the case when the flavors of the neutrino and 
of the charged lepton are different
$(\nu_\mu \to \nu_\mu e^- e^+, \, \dots)$.
 
\newpage
\thispagestyle{empty}

\begin{figure}[ht]
\epsffile[122 270 483 701]{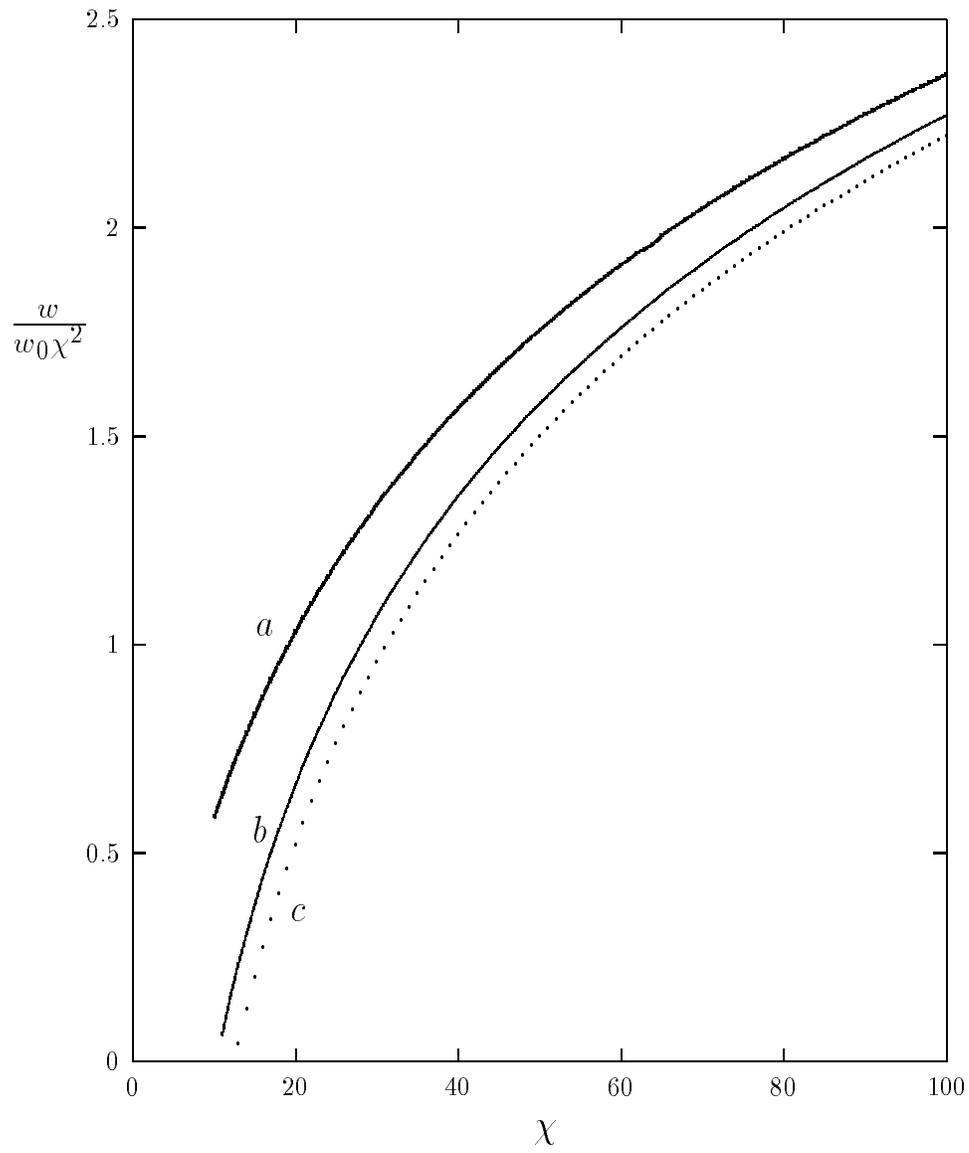}
\caption{A.V.~Kuznetsov et al., ``Lepton pair production \dots''}
\end{figure}

\newpage
\thispagestyle{empty}

\begin{figure}[ht]
\epsffile[122 270 483 701]{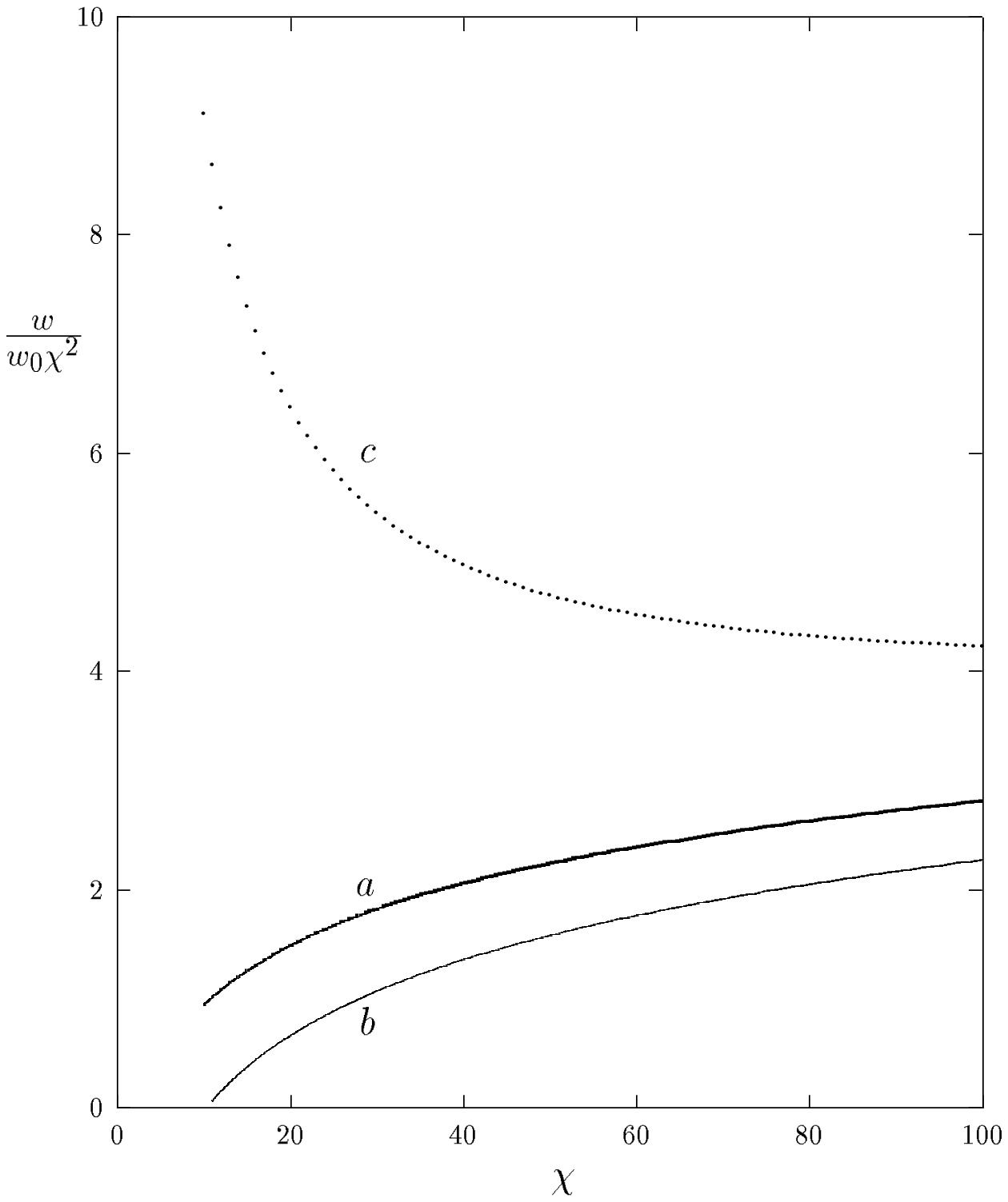}
\caption{A.V.~Kuznetsov et al., ``Lepton pair production \dots''}
\end{figure}


\begin{thebibliography}{9}
\bibitem{Chob}
   E.A.~Choban, A.N.~Ivanov, Zh. Eksp. Teor. Fiz. 56 (1969) 194.

\bibitem{Bor83}
   A.V.~Borisov, V.Ch.~Zhukovskii, B.A.~Lysov, Izv. Vuz. Fiz. 8 (1983) 30 
   [Sov. Phys. J. 26 (1983) 701].

\bibitem{Knizh}
   M.Yu.~Knizhnikov, A.V.~Tatarintsev, Vestn. MGU, ser. Fiz., Astron. 
   25, N 3 (1984) 26.

\bibitem{Bor93}
   A.V.~Borisov, A.I.~Ternov, V.Ch.~Zhukovsky, Phys. Lett. B 318 (1993) 489. 

\bibitem{KM97}
   A.V.~Kuznetsov, N.V.~Mikheev, Phys. Lett. B 394 (1997) 123; 
   Yad. Fiz. 60 (1997) 2038 [Phys. At. Nucl. 60 (1997) 1865].

\bibitem{Skob95}
   V.V.~Skobelev, Zh. Eksp. Teor. Fiz. 108 (1995) 3 [JETP 81 (1995) 1].

\bibitem{Bor99}
   A.V.~Borisov, N.B.~Zamorin, Yad. Fiz. 62 (1999) 1647.

\end{thebibliography}
\end{document}